\documentclass{aa}
\usepackage{graphics}

\def\ltsima{$\; \buildrel < \over \sim \;$}
\def\simlt{\lower.5ex\hbox{\ltsima}}
\def\gtsima{$\; \buildrel > \over \sim \;$}
\def\simgt{\lower.5ex\hbox{\gtsima}}

\def\kms{\ifmmode {\rm \ km \ s^{-1}}  
\else
$\rm km \ s^{-1}$\fi}

\def\Msun{\mbox{$M_\odot$}}

\def\h2{\mbox{\ion{H}{ii}}}

\newcommand{\lessim}{\mathrel{\hbox{\rlap{\hbox{\lower4pt\hbox{$\sim$}}}\hbox{$<$}}}}

\begin{document}
\addtolength{\voffset}{1cm}


\title{Analysis of colour-magnitude diagrams of rich LMC clusters: NGC\,1831}

\author{L. O. Kerber \inst{1}
\and B. X. Santiago \inst{1}
\and R. Castro \inst{1} 
\and D. Valls-Gabaud \inst{2}}

\offprints{kerber@if.ufrgs.br}
 
\institute{Universidade Federal do Rio Grande do Sul, IF, 
CP\,15051, Porto Alegre 91501--970, RS, Brazil
\and UMR CNRS 5572, Observatoire Midi-Pyr\'en\'ees, 14, 
avenue \'Edouard Belin, 31400 Toulouse, France}

\date{Received 08 April 2002 / Accepted 06 May 2002}

\titlerunning{CMD analysis for NGC\,1831}
\authorrunning{Kerber et al.}

\abstract{
We present the analysis of a deep colour-magnitude diagram (CMD) of NGC\,1831, 
a rich star cluster in the LMC. The data were obtained with HST/WFPC2 in the 
F555W ($\sim$ V) and F814W ($\sim$ I) filters, reaching $m_{555} \sim 25$. 
We discuss and apply a method of correcting the CMD for sampling 
incompleteness and field star contamination. 
Efficient use of the CMD data was made by means of direct comparisons of 
the observed to model CMDs.
The model CMDs are built by an algorithm that generates artificial
stars from a single stellar population, characterized by
an age, a metallicity, a distance, a reddening value, 
a present day mass function and a fraction
of unresolved binaries. 
Photometric uncertainties are empirically determined from the data and
incorporated into the models as well.
Statistical techniques are presented and applied as an objective 
method to assess the compatibility between the model and data CMDs.  
By modelling the CMD of the central region in NGC\,1831 we infer
a metallicity $Z = 0.012$, $8.75 \leq \log(\tau) \leq 8.80$, 
$18.54 \leq  (m-M)_{0} \leq 18.68$ and 
$0.00 \leq E(B-V) \leq 0.03$. 
For the position dependent PDMF slope 
($ \alpha = -\mathrm{d}\log\Phi(M)/\mathrm{d}\log M$), 
we clearly observe the effect of mass segregation in the system: 
for projected distances $R \leq 30$ arcsec, $\alpha \simeq 1.7$, 
whereas $2.2 \leq \alpha \leq 2.5$ in the outer regions of NGC\,1831. 

}

\maketitle

\section{Introduction}
\label{intro}

The Large Magellanic Cloud (LMC) presents three essential characteristics that 
make it an excellent complementary laboratory for studying the
formation and evolution of galaxies and stellar systems in general:
a) it is close to the Galaxy; b) it has markedly different morphological,
chemical and kinematical properties from our Milky Way; c) it
presents a large variety of stellar clusters, displaying distinct 
physical characteristics among themselves and when compared to those in
the  Galaxy (Westerlund 1990). Due to the diversity in ages and 
metallicities, LMC clusters are found at distinct evolutionary
stages (Westerlund 1990, Olszewski et al. 1991). 
The determination of a cluster's present physical properties, such as
density profile, shape, internal velocity distribution and 
its position dependent Present Day Mass Function (PDMF),
provide us with essential links needed to assess the role 
of gravitational dynamics, including effects of mass segregation
and stellar evaporation (Heggie \& Aarseth 1992, Spurzem \& Aarseth 1996, 
de Oliveira et al. 2000). Thus, once these present properties are known, 
modelling techniques like N-body simulations allow us to 
recover the initial conditions under which
clusters formed (Goodwin 1997, Vesperini \& Heggie 1997, Kroupa et al. 2001). 
In this context, the initial mass function (IMF) and its possible universality,
are key pieces in the study of stellar contents of distant galaxies (Kroupa
2001).

Through the analysis of colour-magnitude diagrams (CMDs) one has
a great variety of physical information about a cluster. Isochrone fits
help constraining the system's age, metallicity, distance
and reddening. Furthermore, observational luminosity functions (LFs)
have allowed derivation of the stellar mass function 
(Elson et al. 1995, De Marchi \& Paresce 1995, Santiago et al. 1996, 
Piotto et al. 1997, de Grijs et al. 2002a).
However, the transformation of luminosity into mass 
depends on age and metallicity, the uncertainties in these parameters
being therefore incorporated into the inferred mass function.
Besides, the theoretical uncertainties in the mass-luminosity relation itself,
specially for the low-mass stars (Piotto et al. 1997, Baraffe et al. 1998), 
added to the effect of unresolved binarism, further hampers real mass function 
determination through observational luminosity functions.

From both observational and theoretical points of view, the advent of the
Hubble Space Telescope (HST), associated with the constant improvement in 
theories of stellar interiors, atmospheres and evolution, require
ever more sophisticated methods of CMD analysis.
In this context, computational modelling techniques have allowed a 
wider use of CMDs as tools to constrain 
physical properties of stellar populations and systems.
Model CMDs, along with statistical 
methods of comparing them to observed ones, have been
useful means to investigate the star formation history within
a galaxy (Gallart et al. 1996, Gallart et al. 1999, Hernandez et al. 1999, 
Holtzman et al. 1999, Hernandez et al. 2000)
or to constrain structural 
parameters and the stellar luminosity function in the Milky Way
(Kerber et al. 2001). 

With these issues in mind, our work aims at extracting as much
physical information as possible about rich LMC clusters, by means of objective
comparison of their observed CMDs with artificial ones. 
The idea is to simultaneously infer PDMF slope, 
age, metallicity, distance, reddening
and unresolved binary fraction for each system studied. 
The present work introduces the techniques developed for that purpose and
shows the results for NGC\,1831, one of the richest LMC
clusters for which we have deep HST data.

Previous works are evidence of the large difficulties in extracting
physical parameters for NGC\,1831. Techniques 
based on CMDs from CCD photometry (Mateo 1987,1988; Chiosi 1989; 
Vallenari et al. 1992; Corsi et al. 1994), integrated spectroscopy 
or colours (Bica et al. 1986; Meurer et al. 1990; Cowley \& Hartwick
 1992; Girardi et al. 1995) and spectroscopy of individual giant 
stars (Olszewski et al. 1991) were employed with this aim, 
constraining the 
values of the main parameters:
$8.50 \simlt \log(\tau) \simlt 8.80$ ($300 \simlt \tau \simlt 650$ Myr);
$0.002 \simlt Z \simlt 0.020$ ($-1.00 \simlt \mathrm{[Fe/H]} \simlt 0.01$);
$0.00 \simlt E(B-V) \simlt 0.07$.
For the distance to NGC\,1831, there are not reliable determinations, 
the standard procedure being the adoption of typical values of the
intrinsic distance modulus, $(m-M)_{0}$,  for the LMC centre. 
In this aspect, the most reliable estimate seems to be that
of Panagia et al. 1991, $(m-M)_{0}=18.51$, since it is based on purely
geometrical arguments applied to high quality imaging and spectral data on
supernova SN1987a.
In terms of dynamical evolution for this system, Elson et al. (1987) estimated
$6.5 \simlt \log(t_\mathrm{cross}) \simlt 7.0$ and 
$9.6 \simlt \log(t_\mathrm{rh}) \simlt 10.0$ for the crossing time and two-body 
relaxation time, respectively, the range quoted being due to 
different mass-luminosity relations. Comparing with its estimated age values, 
these results suggest that NGC\,1831 is a system dynamically well mixed, 
but not totally relaxed through stellar encounters. 
Hence, NGC\,1831 is sufficiently old to have suffered 
mass segregation, affecting the PDMF slope at different 
distances to its centre, but perhaps still young enough that 
the initial conditions could be preserved in its outer
regions. Similarly, external effects 
may not have had enough time to affect the cluster dynamics either. 

One of the main objectives of this paper is
to verify the effect of mass segregation in NGC\,1831, 
quantifying the variation in PDMF slope with projected 
distance to the cluster's centre. This determination may yield
strong links to IMF reconstruction efforts based on N-body models. 
The paper is outlined as 
follows: in Sect. 2 we present the data and the methods of accounting 
for sample incompleteness and field star contamination in the observed
CMD; in 
Sect. 3 we present the algorithm used for CMD modelling and 
the statistical tools used for model vs. data comparisons; in 
Sect. 4 we discuss control experiments used for to verify the validity
of the method;
finally, in Sect. 5 we present the results for the NGC\,1831 data,
which are discussed in Sect. 6.

\section{The observed CMD}

We have data taken with the {\it Wide Field and Planetary Camera 2}
(WFPC2) on board HST for 8 rich LMC clusters and nearby fields.
These data are part of the GO7307 project, entitled 
``Formation and Evolution of Rich Star Clusters in the LMC'' (Beaulieu
et al. 1999, Beaulieu et al. 2001, Johnson et al. 2001). 
For each cluster, images were obtained using the F555W (V) and 
F814W (I) filters. Most of the photometry had been previously carried out:
cluster stellar LFs were built and analyzed by Santiago et al. (2001),
de Grijs et al. (2002b); 
field stellar populations were studied by Castro et al. (2001).
Exposure times, field coordinates, image reduction and photometry processes
are described in detail by those authors.

\begin{figure}
\resizebox{\hsize}{!}{\includegraphics{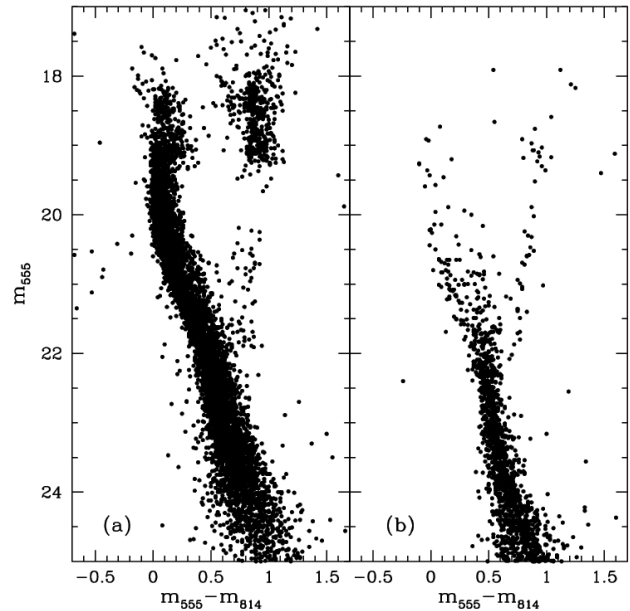}}
\caption[]{The on-cluster (a) and off-cluster (b) WFPC2 CMDs for NGC\,1831. 
The former contains 7801 stars observed in the cluster's direction 
while the latter has 2030 stars located in a field 7.3 arcmin away 
from the cluster's centre.}
\label{cmd_all}

\end{figure}

Fig. \ref{cmd_all} shows the observed WFPC2 CMD for stars in the
direction of NGC\,1831 in panel (a) (hereafter the on-cluster field) and 
for a nearby field (hereafter the off-cluster field)
studied by Castro et al. (2001) in panel (b).
The on-cluster sample shown here is the final composition of the
CEN and HALF images described by Santiago et al. (2001). These have
the Planetary Camera (PC) centred on the cluster's centre and half-light
radius, respectively.
The off-cluster field is located at about 7.3 arcmin away from the 
cluster's centre. 
A clear cluster main sequence (MS) is visible in the figure, stretching from
$m_{555} \simeq 18.5$ down to $m_{555} \simeq 25$. The cluster 
MS turn-off is also clearly visible at the upper MS end. 
Notice, however, that saturation becomes a problem in the HALF field for
$m_{555} \simlt 19$ ($m_{555} \simlt 17.8$ for the CEN field). Hence, all
our subsequent analysis will be based on the CMD fainter than this limit.
A branch of evolved 
stars is seen as well, especially in the range  $18 \simlt m_{555}
\simlt 19$, where the cluster red clump is located. The
subgiant branch at fainter magnitudes is due to field contamination and is 
largely made up of older ($\tau > 3$ Gyr) stars.

The on-cluster data suffer from two
important effects: sample incompleteness and contamination by field stars. 
Our CMD modelling algorithm does not incorporate such
effects. Therefore it is crucial to adequately correct the observed
CMD for them in order to place models and data on equal footing.
Quantifying systematic and random photometric uncertainties 
and either correcting for them or applying them
to model CMDs is extremely important as well, as they are 
responsible for most of the observed CMD spread.
These data corrections are the subject of the next subsections.

\subsection{Random photometric uncertainties}
The random photometric uncertainties are the main source of spread
in our HST/WFPC2 CMDs. 
Therefore, a suitable assessment of these uncertainties in both 
filters is crucial for correctly incorporating this effect into 
the artificial CMDs.

For the on-cluster field, two independent photometric measurements 
were available for a fraction of the stars due to the overlap region 
imaged by both the HALF and CEN fields (Santiago et al. 2001).
Thus, we used the stars belonging to this overlap region to estimate
the typical photometric uncertainties in the data. 
For each filter and at each magnitude bin, we estimated the
dispersion, $\sigma'$, in the distribution of differences between
the independent magnitude measurements. For simplicity we assume that 
$\sigma'$ is the composition of two equally-sized realizations
of photometric error, $\sigma$.
Thus, we get $\sigma'^2 = 2\sigma^2$, and therefore

$$\sigma =\frac{\sigma'}{\sqrt{2}} .$$

We emphasize that the two filters were treated as absolutely independent. 
As a consequence, the uncertainty in colour $m_{555}-m_{814}$, 
$\sigma_\mathrm{colour}$, will be the quadratic sum of the individual uncertainties:

$$\sigma_\mathrm{colour}^2 = \sigma_{555}^2 + \sigma_{814}^2 .$$

\begin{figure} 
\resizebox{\hsize}{!}{\includegraphics{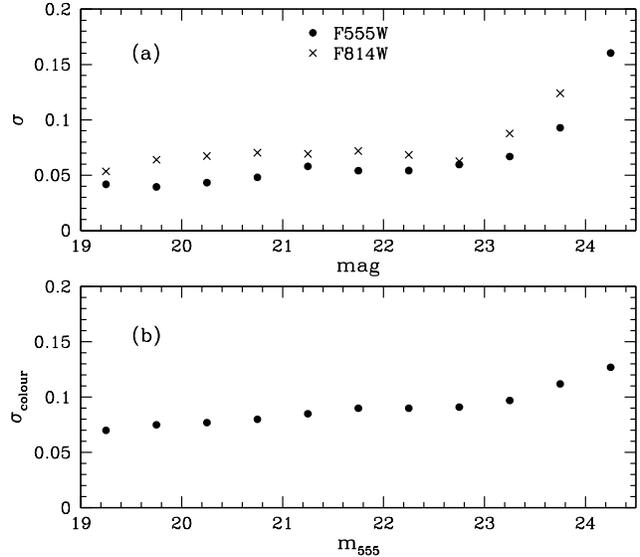}}
\caption[]{Estimated photometric uncertainties. Panel (a) shows the 
$\sigma_{555}$ (solid circles) and $\sigma_{814}$ (crosses) values 
as a function of apparent magnitude. 
Panel (b) shows the $\sigma_\mathrm{colour}$ calculated for the MS stars 
as a function of $m_{555}$.}
\label{error}
\end{figure}

Fig. \ref{error} shows the derived uncertainties for MS stars in 
the two filters, 
$\sigma_{555}$ and $\sigma_{814}$, and for the colour, $\sigma_\mathrm{colour}$, 
using the expressions above.
 
For the off-clusters stars, we did not have two independent and
overlapping WFPC2 images and, as
a result, we could not apply the same method to quantify their photometric
uncertainties. The solution found was to employ the same estimate
as for the on-cluster stars. As the off-cluster image is deeper 
and sparser than the on-cluster ones, 
we can expect that this approximation yields an overestimate of 
the photometric uncertainties in the off-cluster data. 
However, the off-cluster CMD is used only for statistical subtraction of
field contamination from the on-cluster CMD. We will see later that this
particular correction technique is fairly independent of the 
estimated photometric uncertainties.

\subsection{Systematic photometric uncertainties}

Systematic effects in WFPC2 data have been found by several authors.
Johnson et al. (2001) measured an exposure time effect varying
from 0.01 to 0.06 for different WFPC2 chips and filters
in images of NGC\,1805 and NGC\,1818, as part of this project. 
de Grijs et al. (2002b) finds similar trends, but
of slightly larger amplitude for the same data. Previously,
Casertano \& Muchtler (1998) found an exposure time effect 
but in the opposite sense as Johnson et al. (2001).
Colour shifts of $\simeq 0.04$ have also been measured between different 
WFPC2 chips, possibly
due to errors in CTE and aperture corrections or in zeropoints (see also
Johnson et al. 1999).

Any photometric biases, as a function of either exposure time or chip, 
must be eliminated from our observed CMD,
since the model CMDs which will be compared to it do not
incorporate them.

\begin{figure} 
\resizebox{\hsize}{!}{\includegraphics{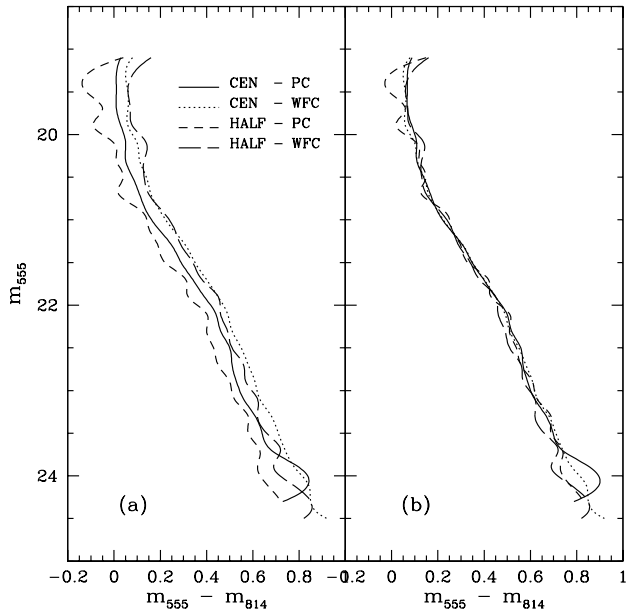}}
\caption[]{MS fiducial lines for each chip (PC or WFC) and
image (CEN or HALF) as indicated: (a) original MS lines; (b) MS lines 
after applying
the correction discussed in the text.}
\label{fidlines}
\end{figure}

As there does not seem to exist a consensus on the corrections to
be applied, our approach was to empirically measure such biases
and to apply appropriate shifts to the data when necessary.
We searched for both exposure time and chip vs. chip effects.
No systematic effects were found in the off cluster data.
The main source of bias in the on-cluster data was found to be
an offset between the PC and the Wide Field Camera (WFC) chips
in the sense that
MS stars with the same $m_{555}$ magnitude tend to be bluer 
by 0.05-0.10 mag when imaged with the PC than with the WFC. This applies
to both HALF and CEN images.
As the PC in the HALF image is centered on the cluster half-light
radius, this effect is unlikely to be due to differences in 
crowding. In order to correct the data for this effect we first defined
MS fiducial lines, taking the median $m_{555}-m_{814}$ 
colour at different magnitude bins. This was done separately for
each chip and each image (CEN or HALF). We noticed that the WFC MS lines were
more stable, always occupying loci in the CMD very close to each other.
Thus, we transposed the PC fiducial lines to the 
corresponding WFC locus. The corrections are shown in
Fig. \ref{fidlines}. Panel (a) shows the uncorrected PC and WFC (a mean
locus of the 3 chips) lines for both the CEN and HALF images. Panel (b)
shows the corrected fiducial lines.

\subsection{Selection effects}

\subsubsection{Sample incompleteness}
Sample incompleteness occurs essentially due to two factors: overlapping
of stellar profiles due to crowding and background noise. 
Therefore, in a cluster incompleteness will depend not
only on apparent magnitude, but also on spatial position. Faint stars
in dense stellar regions are the ones that suffer most from
this effect. 
Completeness effects in the on-cluster data 
were previously discussed and measured 
by Santiago et al. (2001). These authors carried out 
experiments where artificial stars were added to
the cluster image and subjected to the same sample
selection as the real stars. 
The completeness {\it c} of each real star was estimated
as the fraction of artificial stars of similar
magnitude and location which were successfully recovered in the
experiments. The estimated weight 
{\it w} for each star is then simply given by the inverse of 
its completeness ($w = 1/c$).
This weight corresponds to the total number of stars similar to the
observed one which should be detected and measured in an ideal image.

Assigning a weight to each observed star is enough for the sake of luminosity
and mass functions. However, correcting a CMD for incompleteness requires
an extra step, namely 
to place an extra number of stars on the CMD according to the position
of each observed star and its previously computed completeness weight.
In order to fill the observed CMD with the missing stars, we first
defined a fiducial line for the MS. As described in Sect. 2.2,
this line was defined by taking the median $m_{555}-m_{814}$ 
colour at different magnitude bins. 
Given the $i^{th}$ MS star, its position along the fiducial line is 
provided by its $m_{555}$ magnitude and the corresponding $m_{555}-m_{814}$ 
colour. If its completeness weight is $w_i$, 
then $w_i - 1$ extra stars were 
spread out from its position along the MS line taking the 
measured photometric uncertainties (as described in Sect. 2.1) into account.
We assumed a Gaussian distribution of uncertainties both in
$m_{555}$ magnitude and $m_{555}-m_{814}$ colour.
Sample completeness falls to less than $50\%$ for
$m_{555} \simgt 23.8$. We completed the CMD down to $m_{555}=24.5$
and then cut it at $m_{555}=23.5$, therefore avoiding boundary effects.
A complete sample of stars 
both in number and in position resulted from this method.

As for the off-cluster CMD, it is complete at least down to $m_{555} =
24.5$ (Castro et al. 2001).

\subsubsection{Field star subtraction}
Field star subtraction from the cluster sample
is carried out in two separate steps.
We first cut-off all stars in the on-cluster CMD which are much farther
from the MS than expected by photometric errors. We therefore eliminate
all evolved stars, as well as objects which are likely to be foreground
stars or remaining non-stellar sources in the sample 
(distant galaxies, spurious image features, etc).

The second step involves the statistical removal of field stars
located along the MS. We thus
compare the distribution of stars in the on-cluster CMD to 
that of the off-cluster CMD.
The comparison method is based on the hypothesis that the positions 
of the off-cluster
stars represent the most likely positions for field stars on any
similar CMD. We thus try to estimate the probability of each on-cluster
star to be a field star and according to this probability we randomly
remove stars from the on-cluster CMD.

We consider pairs of on-cluster/off-cluster stars. 
For each pair we compute
the expected scatter in CMD position between the pair members under
the assumption that they are independent photometric realizations of the
off-cluster star. We will then have

$$\sigma_\mathrm{c,555} = \sqrt{2} \sigma_\mathrm{off,555}$$

and 

$$\sigma_\mathrm{c,colour} = \sqrt{2} \sigma_\mathrm{off,colour}$$

respectively for the expected scatter in $m_{555}$ magnitude and 
$m_{555} - m_{814}$ colour, where the ``off'' subscript in the
expressions above refers to the off-cluster pair member.

For the $i^{th}$ off-cluster CMD star we then consider the $N_{i}$ 
on-cluster stars inside a 
$3\sigma_\mathrm{c,555}$ x $3\sigma_\mathrm{c,colour}$
box centered on it.
Using a Gaussian error distribution in magnitude and colour, we 
estimate the probability $p_{i,j}$ that the $j^{th}$ on-cluster CMD star,
inside this box, is a second photometric measurement of the 
$i^{th}$ field star.  Therefore, 

$$p_{i,j} \propto exp{[\frac{-(mag_i-mag_j)^2}{2(\sigma_\mathrm{c,555})^2}]}
exp{[\frac{-(colour_i-colour_j)^2}{2(\sigma_\mathrm{c,colour})^2}]}$$

where the normalization of $p_{i,j}$ is such that

$$\sum_{j=1}^{N_i}p_{i,j} = 1$$

Doing the same for all $N_\mathrm{off}$ off-cluster CMD stars, we estimate
the probability $P_{j}$ that the $j^{th}$ on-cluster CMD star is any one of
the $N_\mathrm{off}$ field stars. Hence, 

$$P_j = \sum_{i=1}^{N_\mathrm{off}} p_{i,j} ,$$ where 
$$\sum_{j=1}^{N_\mathrm{on}} P_j = N_\mathrm{off}$$ and 
$N_\mathrm{on}$ is the total number of stars in the on-cluster CMD.
In practice, not all off-cluster stars will have on-cluster stars
within their 
$3\sigma_\mathrm{c,555}$ x $3\sigma_\mathrm{c,colour}$ boxes. 
The actual off-cluster stars taken into account will therefore be 
$N_\mathrm{off}'< N_\mathrm{off}$.

Based on the $P_j$ probabilities, and scaling the number of field stars
to the solid angle of the on-cluster field,
we randomly remove

$$N_\mathrm{field} = N_\mathrm{off}' \frac{\Omega_\mathrm{on}}
{\Omega_\mathrm{off}}$$

stars from the on-cluster CMD, where $\Omega_\mathrm{on}$ and 
$\Omega_\mathrm{off}$ are
the solid angles covered by the on-cluster and off-cluster fields, 
respectively.

\begin{figure} 
\resizebox{\hsize}{!}{\includegraphics{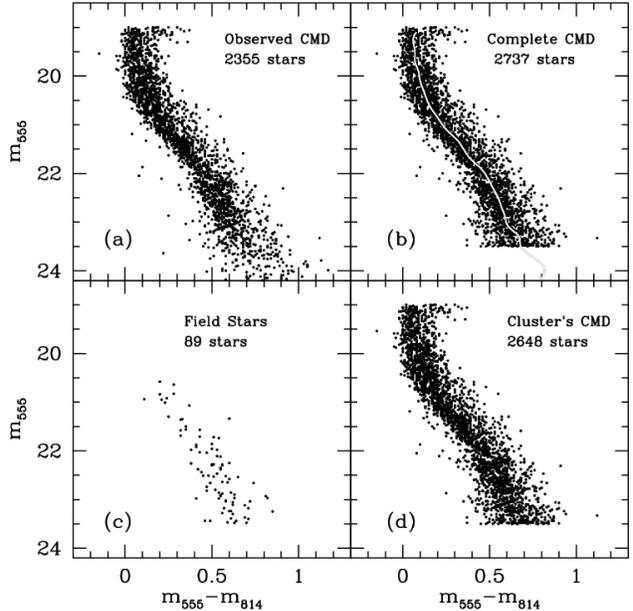}}
\caption[]{Correcting the on-cluster CMD for incompleteness and field stars
contamination:
(a) the observed CMD along the MS; 
(b) the CMD corrected for sample incompleteness and cut at $m_{555}$=23.5.
The MS fiducial line for the cluster is shown in white ;
(c) on-cluster stars considered as LMC field stars according
to the algorithm described in the text;
(d) final CMD with only stars presumed to belong to the cluster.}
\label{4cmd_central}
\end{figure}

Fig. \ref{4cmd_central} shows the results of correcting an observed
NGC\,1831 CMD for incompleteness and field contamination. 
Panel (a) shows the original CMD 
obtained from photometry, excluded of non-MS stars and 
corrected only for the systematic photometric
effects described in Sect. 2.2.
Panel (b) presents the complete CMD, i.e., corrected for
sample incompleteness and cut at $m_{555} = 23.5$; 
the extra stars added by completeness correction represent about 23\% of
the total within this magnitude range. 
Panel (c) shows the 89 stars ($\simeq$ 3\%) in the on-cluster CMD
that were considered as LMC field stars in the field
star subtraction process; finally the 
clean and final cluster CMD is shown in panel (d).

We tested the field star subtraction algorithm for different box sizes and 
assumptions regarding the photometric scatter. The results are insensitive to
the details in the algorithm.

\section{CMD modelling and statistical tools}

\subsection{CMD Modelling}
We model the MS of a cluster as a single stellar 
population, characterized by stars of the same age and chemical 
composition. The first step is the choice of 
an isochrone, which defines a sequence of magnitude and colour as a function
of mass for this population.
For the present work we used Padova isochrones (Girardi et al. 2000)
because they present masses inside the observed MS mass range
($M \simlt 2.5$ \Msun) and are expressed in the vegamag WFPC2 
photometric system. Notice that Padova models assume convective 
overshooting in the stellar interiors. This assumption may influence
age determinations, specially when based in the position of turn-off
and He-burning stars of younger populations (Testa et al. 1999, 
Barmina et al. 2002).  However, as NGC\,1831 is at least several Myr old
and our modelling makes use of the entire MS,
we believe that these model uncertainties tend to be of smaller 
importance in our analysis.

The free model parameters are: metallicity ($Z$), age ($\log(\tau)$),
intrinsic distance modulus $(m-M)_{0}$, reddening (E(B-V)), PDMF slope 
($\alpha=-\mathrm{d}\log\Phi(M)/\mathrm{d}\log M$) 
and unresolved binary fraction
 ($f_\mathrm{bin}$). The PDMF was considered as a power-law 
($\Phi(M)=\Phi_{0}M^{-\alpha}=\mathrm{d}N/\mathrm{d}M$), 
where the only free parameter is
the slope $\alpha$. $f_\mathrm{bin}$ is defined as the systemic 
binary fraction,
$f_\mathrm{bin}=N_\mathrm{bin}/(N_\mathrm{bin}+N_\mathrm{sing})$,
where $N_\mathrm{bin}$ and $N_\mathrm{sing}$ are, respectively, the number of 
unresolved pairs and single stars.

The process of generating artificial stars works as follows: 

\smallskip

(1) we fix $Z$ and $\log(\tau)$ for the stellar population by means of a
chosen isochrone;

\smallskip

(2) we randomly draw a stellar mass according to the PDMF and get the absolute
magnitudes in the two desired filters through the mass-luminosity 
relation given by the isochrone; 

\smallskip

(3) for $f_\mathrm{bin}$ randomly chosen cases, we repeat step (2), representing a 
companion star in a binary system, and combine the two luminosities in 
both filters;

\smallskip

(4) we apply the intrinsic distance modulus $(m-M)_{0}$ and reddening 
vector($A_\mathrm{V},E(B-V)$) to the system, defining its theoretical 
CMD position. 
For this purpose, we use $R_\mathrm{V}=A_\mathrm{V}/E(B-V)=3.1$ and 
the photometric 
transformation to the vegamag WFPC2 system according to Holtzman et al. 1995a;

\smallskip

(5) we introduce the photometric uncertainties by spreading the star with
a Gaussian distribution of errors with $\sigma_{555}$ and $\sigma_{814}$
as empirically determined (see Sect. 2.1).
This yields {\it observational versions} of the magnitude and
colour;

\smallskip

(6) finally, we verify if the star is inside the MS observational ranges
in magnitude and colour defined for the data and throw it away if it is not.

\smallskip

For each model realization we generate the same number of MS stars as
observed in the real CMD, corrected for sample incompleteness and field
contamination, and located inside the $19.0 \leq m_{555} \leq 23.5$ range. 
This range in apparent magnitude corresponds to 
$0.5 \simlt M_{555} \simlt 5.0$ and $0.9 \simlt M \simlt 2.3$ \Msun.

The best models are chosen by a direct comparison of the observed CMD with
the artificial ones. The statistical tools used in this comparison are
presented in the next section. The model vs. observation
comparison strategy is as follows: 

\smallskip

(1) the global parameters for the cluster, $\log(\tau)$, $Z$, $E(B-V)$ and
$(m-M)_{0}$, are determined using the CMD of the central cluster region
($R \leq 30$ arcsec, where $R$ is the projected distance from the cluster's
centre), where field contamination and statistical noise 
are minimized (see Table \ref{tab1});

\smallskip

(2) for the best combinations of the global parameters,
the position dependent parameters $\alpha$ and $f_\mathrm{bin}$
are then derived using the CMDs in concentric rings of variable radii.

\smallskip

Table \ref{tab1} shows important information about the cluster regions used
in this modelling process. The first column gives the range in $R$, Col.
(2) the original number of CMD stars in the $19.0 \leq m_{555} \leq 23.5$
range, Col. (3) the number of stars
in the completeness corrected
CMD (see Sect. 2.3.1), Col. (4) the number of assigned field 
stars (Sect. 2.3.2) and, finally,
Col. (5) the number of stars in the CMD used in the modelling process.
Notice that, not unexpectedly, field contamination 
becomes a serious issue for the outermost region, since statistical removal
of field stars reduces the CMD numbers by about 40\%.
On the other hand, crowding in the central regions yields larger 
photometric incompleteness, as reflected by the clear increase in the
number of stars between Cols. (2) and (3).

This modelling strategy allows an efficient and systematic search 
of best fit models
in a 6-dimensional parameter space and makes use of the 2-dimensional 
information provided by the CMD data. Furthermore, this strategy naturally
splits the parameters into those that define the position of the MS in the CMD
plane (the global ones) and those that influence the way stars are distributed
within the MS locus (the position dependent ones).

\begin{figure} 
\resizebox{\hsize}{!}{\includegraphics{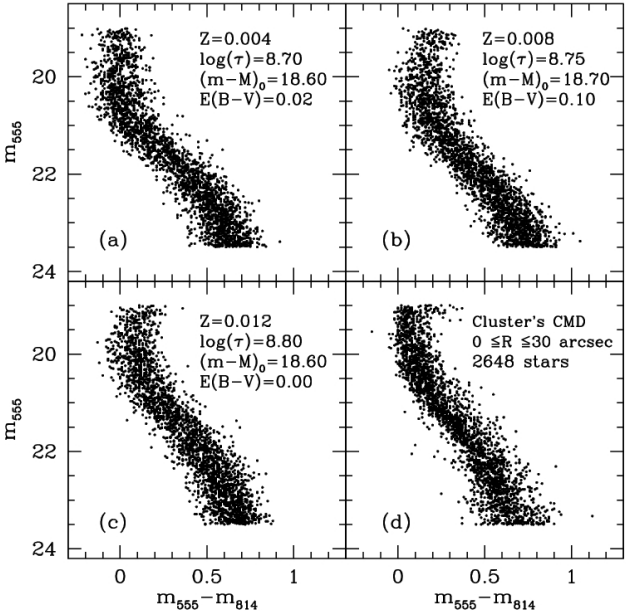}}
\caption[]{CMDs for the cluster central region: panels a, b and c show 
realizations from the modelling process while panel d shows the result 
of the real data treatment (Sect. 2). 
The input model parameters are indicated in each 
panel, where $\alpha=2.30$ and $f_\mathrm{bin}=0.30$ for all models.}
\label{modelcmds1}
\end{figure}

Fig. \ref{modelcmds1} shows four CMDs for the central cluster region. 
The one in the bottom right (panel d) is the real data, whereas the other 
three are realizations from different models generated by the modelling 
process described above. 
The input model parameters are shown in each panel. The three model CMDs
do in fact look different, their MS ridge lines having different shapes and
occupying different positions along the CMD plane. 

\begin{table}
\caption[]{Number of points in the different cluster regions whose CMDs
are used in the CMD modelling as described in Sect. 3.1.}
\label{tab1}
\small
\renewcommand{\tabcolsep}{1.1mm}
\begin{tabular}{lcccc}
\hline\hline
{Region} (arcsec) & {observed} & {complete} & {field stars} & {cluster}
\\ \hline
\hline
$0 < R \leq 30$ & 2221 & 2737 & 89 & 2648 \cr
$0 < R \leq 15$ & 1220 & 1506 & 27 & 1479 \cr 
$15 < R \leq 30$ & 1001 & 1216 & 62 & 1154 \cr
$30 < R \leq 60$ & 1692 & 1972 & 240 & 1732 \cr
$R > 60$ & 1673 & 1780 & 765 & 1015 \cr
\hline\hline
\end{tabular}
\end{table}

\subsection{Statistical tools}
One of the main goals of this paper is to establish an objective 
comparison method between models and data. This required developing and
applying statistical criteria that discriminate the 
model CMDs that most adequately reproduce the observed one.
Ideally these comparison criteria should be both simple and easy to
implement but yet make use 
of as much information provided by the bidimensional 
colour-magnitude plane as possible. We stress
that these methods, in principle, are not restricted only to CMD analyses, 
but may be applied to the comparison of any two 
bidimensional distributions of points.
In similar context as in this work, statistical techniques 
of CMDs comparison have been successfully applied by
Kerber et al. (2001), Lastennet \& Valls-Gabaud (1999), Saha (1998), 
Hernandez et al. (1999), allowing a model
reconstruction of the main CMD features displayed by the
component stars within a galaxy or a cluster.

Three statistics of CMD comparison were considered in this paper:
$S^{2}$, $PSS$ and $L$. The first two are explained in more detail 
by Kerber et al. (2001). $S^{2}$ is essentially a dispersion between
model and data points in the CMD plane; 
$PSS$ is proportional to the joint probability 
that the two CMDs being compared are 
drawn from the same population.

As for the $L$ statistics, it is an empirical version of the 
likelihood statistics described and used by 
Hernandez et al. (1999). 
For each model, 300 realizations with the same number of artificial stars
as the real data (hereafter $N_\mathrm{obs}$) were generated. Dividing 
the CMD plane into $N_\mathrm{b}$ boxes,
the model probability of one star, randomly chosen
from any of these 300 realizations, 
to be inside the $k^{th}$ box is given by 
$p_{k}=N_{k}/(300~N_\mathrm{obs})$. $N_{k}$ is the sum of stars from all
realizations which fall in the $k^{th}$ box.
Thus, $p_k$ may be interpreted as a probability function 
along the CMD plane.

The likelihood  L of a given model is then defined as

$$ L = \prod_{i=1}^{N_\mathrm{obs}} p_{\mathrm{obs},i}$$ 

or 

$$\log L = \sum_{i=1}^{N_\mathrm{obs}} \log(p_{\mathrm{obs},i}),$$ 

where the product and sum are over the $N_\mathrm{obs}$ observed stars, and
$p_{\mathrm{obs},i}$ is the model probability function evaluated at the CMD position
of the $i^{th}$ observed star.

So, for each model we have three distinct statistical values. 
In order to establish the best models we build diagnostic diagrams 
(hereafter DDs), 
which are planes where we confront any two of these statistics.
The best models will naturally have large $L$ and $PSS$, 
and small $S^2$ values.
The method was tested by means of control experiments,
that are shown in Sect. 4.

\subsection{The model grids}

The model input values for $\log(\tau)$, $Z$, $(m-M)_{0}$, $E(B-V)$ 
were chosen in order to bracket those found in the literature. 
In this respect, the web
page www.ast.cam.ac.uk/STELLARPOPS/LMCdatabase by Richard
de Grijs was very useful as it includes a very large
compilation of parameter values and references on
the 8 LMC rich clusters imaged by the GO7307 project.

In accordance with this database, we set the range of possible
values for each physical parameter and created a regular model grid within
this range. We expect this systematic grid to prevent biases in the
parameter values determination. 

Using the cluster central region ($0 \leq R \leq 30$ arcsec), 
we explored the following space defined
by the global parameters:

\smallskip
$Z=0.004, 0.008, 0.012$

\smallskip
$\log(\tau /\mathrm{yr})=8.70, 8.75, 8.80$

\smallskip
$(m-M)_{0}=18.30, 18.40, 18.50, 18.60, 18.70$

\smallskip
$E(B-V)=0.00, 0.02, 0.04, 0.06, 0.08, 0.10$

\smallskip
The position dependent parameters were kept fixed 
as $\alpha = 2.30$ and $f_\mathrm{bin} = 0.30$.
Therefore this initial grid has 270 models. 

Once the CMD comparison statistics defined in Sect. 3.2 are computed
for each model, the DDs are built and the best models are identified, 
we investigate
the dynamically affected, position dependent parameters 
by generating artificial CMDs to be compared to the observed CMDs
within the concentric rings. In this second step we explored
the following bidimensional parameter space: 

\smallskip

$\alpha=1.40, 1.70, 2.00, 2.30, 2.60$

\smallskip

$f_\mathrm{bin}=0.25, 0.50, 0.75$

\smallskip

A total of 180 models were built for each ring, the
only difference between one ring and another being the number of 
artificial stars. 

\section{Control Experiments}

As mentioned in Sect. 3.2, we tested the validity of
our statistical methods using control experiments. 
These experiments consist of drawing one 
realization of some specific model and calling it the
``observed CMD''. 
We then verify if the DDs recover the generating model
(hereafter input model) as the best one describing the ``observed 
CMD''.

\begin{figure} 
\resizebox{\hsize}{!}{\includegraphics{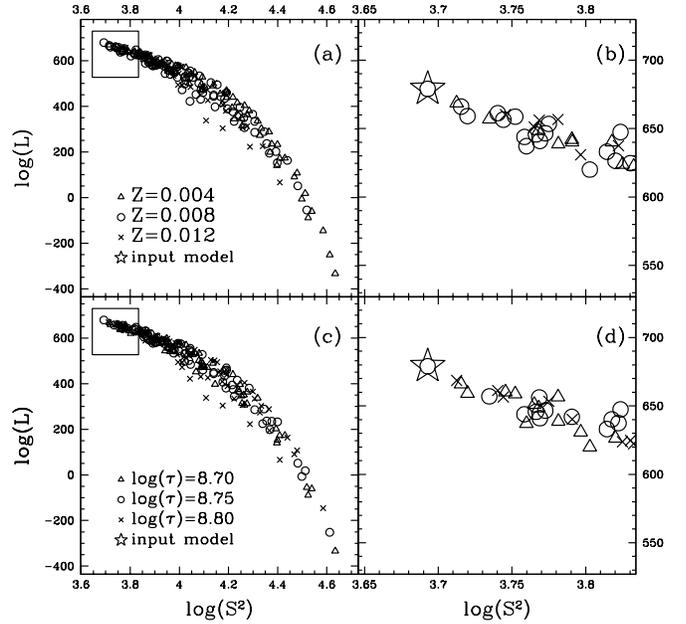}}
\caption[]{DDs resulting from the control experiment
for the cluster's central region,
showing the effects of varying metallicity (panels (a) and (b)) and age
(panels (c) and (d)). The symbols are as indicated in the panels on the left. 
The panels on the right show the best models in detail and use the same symbol 
notation.}
\label{ddfkcen1}
\end{figure}

\begin{figure} 
\resizebox{\hsize}{!}{\includegraphics{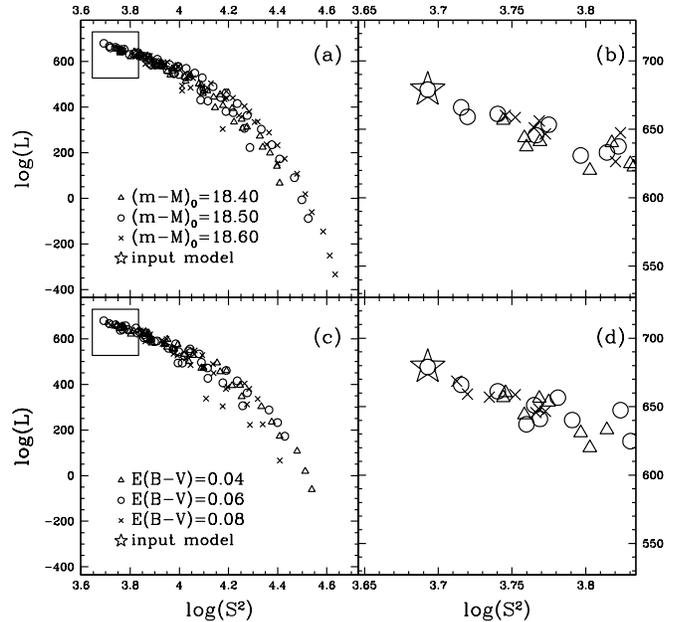}}
\caption[]{DDs resulting from the control experiment for the cluster's central
region, showing the effects of varying distance (panels (a) and (b))
and reddening (panels (c) and (d)). 
The symbols are as indicated in the panels on the left. 
The panels on the right show the best models in detail.}
\label{ddfkcen2}
\end{figure}

Figs. \ref{ddfkcen1} and \ref{ddfkcen2}
show the results of a control experiment involving 
the 270 models to be latter applied to the 
NGC\,1831 central region. All panels show DDs of $\log S^{2}$ vs. $\log L$, 
each point in the DD representing a particular model.
The panels on the right are blow-ups of those on the left, showing in detail
the region where the
best models are located; this region corresponds to larger $\log L$ and
smaller $\log S^2$ values.
The different symbols in a single panel are coded according
to the values of one of the four 
global model parameters ($Z$, $\log(\tau)$, $(m-M)_{0}$ and $E(B-V)$), 
therefore
allowing the effect of varying each parameter to be separately assessed. 
Notice that the figures do not show the entire model grid in order 
to avoid cluttering.
The grid regions discarded from the DDs are those of systematically
high $\log S^2$ and low $\log L$ values.
The ``observed CMD'' was taken to be a
realization of the model with $Z=0.008$, $\log(\tau)=8.75$, $(m-M)_{0}=18.50$
and $E(B-V)=0.06$. This input model is shown as the large star in 
the blowup panels.

A tight correlation between the two statistics is clearly seen in all panels,
adding reliability and stability to the choice of the best fitting models.
The control experiments also show that one is in fact
capable of recovering the input model based on the values
of the statistics used: {\it it is the model with largest $\log L$ and
smallest $\log S^2$, as ideally we would expect.}

Another important result from Figs. \ref{ddfkcen1} and \ref{ddfkcen2}
is the combining and/or canceling effect of some global parameters, yielding
models of comparable quality. As an example, the effects of metallicity
$Z$ and reddening $E(B-V)$ tend to cancel each other. 
Some models with $Z$ ($E(B-V)$)
lower (higher) than the input value, along with some
high $Z$ and low $E(B-V)$ ones, rank among the best models in the DDs.
This degeneracy in the DDs is not surprising
since the effect of increasing $Z$ is to make stars redder and fainter at a
given mass, roughly opposite to the effect of decreasing $E(B-V)$.

\begin{figure} 
\resizebox{\hsize}{!}{\includegraphics{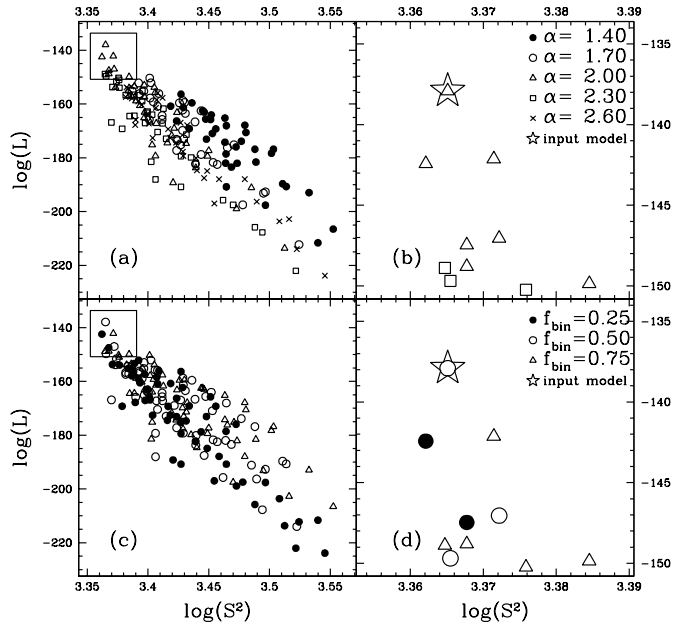}}
\caption[]{DDs resulting from the control experiment for 
the $ 15 \leq R \leq 30$ arcsec region. 
Panels (a) and (b) (the latter is a blowup of the former)
show the effect of varying PDMF slope $\alpha$, whereas
panels (c) and (d) (the latter is a blowup of the former)
show the effect of varying $f_\mathrm{bin}$. 
The input model is represented by the large star.}
\label{ddfkrings}
\end{figure}

Fig. \ref{ddfkrings}
presents similar DDs as in Figs. \ref{ddfkcen1} and \ref{ddfkcen2},
but shows the model grid to be applied to the concentric regions of NGC\,1831 
(in a total of 180 models). 
The symbols now indicate different values of the PDMF slope $\alpha$
(panels (a) and (b)) and binary fraction $f_\mathrm{bin}$ (panels (c) and (d)).
As before, the panels on the right are blow-ups of the ones on the left,
showing the best models only. The input model (large star), in this case, 
is the one with $\alpha=2.00$, $f_\mathrm{bin}=0.50$ (and
$Z=0.012$, $\log(\tau)=8.75$, $(m-M)_{0}=18.60$, $E(B-V)=0.02$). 
It is again located at the extreme upper left corner of the DD, 
confirming the applicability of our statistical approach. 

However, the panels on the right show that the three best models have 
the same $\alpha$ ($= 2.00$) 
but different $f_\mathrm{bin}$ values, revealing a larger 
difficulty in determining the latter than the former.
This occurs because the effect caused by binaries 
is of the same order as or smaller than 
the photometric uncertainties in our WFPC2 CMDs. 
Consequently, as will be discussed in Sect. 5.2, the 
$f_\mathrm{bin}$ determination by means of our CMDs becomes a difficult task.

Notice that, in comparison
with the DDs for the central region, the DDs in Fig. \ref{ddfkrings} 
present a larger dispersion. 
This is caused by the much smaller number of 
stars used in the set of models with
varying $\alpha$ and $f_\mathrm{bin}$; in this case, all model CMDs
(including the ``observed'' one) have 1154 stars, therefore mimicking the 
situation of the second
concentric region to be studied in NGC\,1831 (see Table \ref{tab1}).

\begin{figure} 
\resizebox{\hsize}{!}{\includegraphics{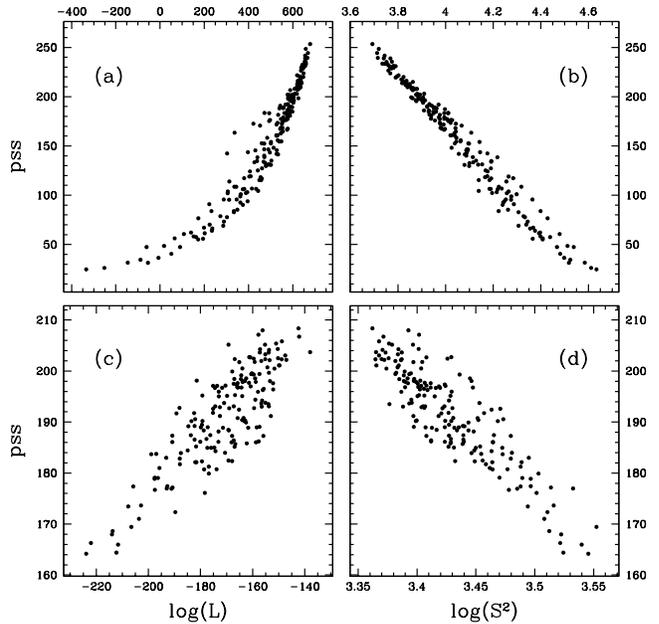}}
\caption[]{DDs involving the PSS statistics in control experiments. 
The upper (lower) panels show $PSS$ vs. $\log L$ and $PSS$ vs. $\log S^{2}$ 
plots for the region inside $0 \leq R \leq 30$ ($ 15 < R \leq 30$) arcsec.}
\label{ddfkpss}
\end{figure}

Fig. \ref{ddfkpss} shows DDs involving the $PSS$ statistics.
The upper panels show $PSS$ vs. $\log L$ and $PSS$ vs. $\log S^{2}$ plots
for the central region. The lower panels show the same plots
for one of the concentric rings. The same models as in Figs.
\ref{ddfkcen1}, \ref{ddfkcen2}, and \ref{ddfkrings}
are depicted but without symbol coding as a function of parameter values.
The correlation between $PSS$ and the other statistics is again quite tight.
In fact, the results based on the DDs are insensitive to the particular
choice of statistics to be plotted. This is a very important result, since
it further enhances the reliability of our statistical CMD modelling 
techniques. 
For the sake of simplicity, we hereafter restrict ourselves to 
$\log L$ vs. $\log S^{2}$ DDs only. 

\section{Results}

\subsection{Central region}
In this section we model the CMD of stars belonging 
to the region within $R \leq 30$ arcsec from the centre of NGC\,1831.
In accordance with Santiago et al. (2001), this corresponds roughly to 
$R \simlt 7$ pc or within 2 half-light radii. 
As indicated in Table \ref{tab1},
there are 2221 stars in this region in the magnitude range
$19.0 \leq m_{555} \leq 23.5$. 
As also mentioned earlier, this central cluster region suffers from
insignificant contamination by LMC field stars.
On the other hand, due the high stellar density, incompleteness 
effects become more important. The chosen faint magnitude cut-off 
represents the magnitude at which 
completeness falls at $50\%$, in an attempt to
reduce the relevance of this effect on our results.

\begin{figure} 
\resizebox{\hsize}{!}{\includegraphics{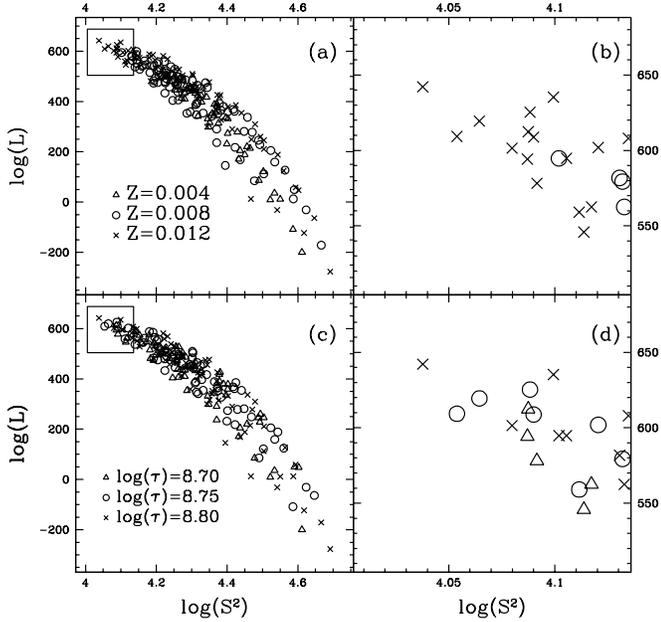}}
\caption[]{DDs resulting from the CMD of the cluster's central region,
showing the effects of varying metallicity (panels (a) and (b)) and 
age (panels (c) and (d)).
The symbols are as indicated in the panels on the left.
The panels on the right show the best models in detail.}
\label{ddobscen1}
\end{figure}

\begin{figure} 
\resizebox{\hsize}{!}{\includegraphics{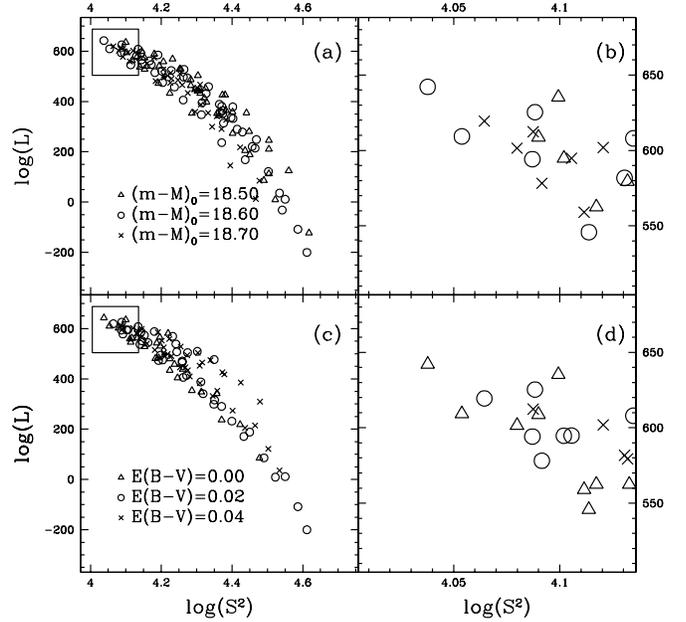}}
\caption[]{DDs resulting from the CMD of the cluster's central region,
showing the effects of varying distance (panels (a) and (b)) and 
reddening (panels (c) and (d)).
The symbols are as indicated in the panels on the left.
The panels on the right show the best models in detail.}
\label{ddobscen2}
\end{figure}

The DDs for the model vs. real CMD comparison are presented in 
Figs. \ref{ddobscen1} and \ref{ddobscen2}. 
As in the control experiments, the correlation between $\log L$ and
$\log S^2$ is quite noticeable, 
the best models being again at the upper left region in the panels.
These figures follow the same
conventions and notations as Figs. \ref{ddfkcen1} and \ref{ddfkcen2},
therefore allowing us to assess the effect of varying each global cluster
parameter separately. 
For example, panels \ref{ddobscen1}(a) and \ref{ddobscen1}(b) clearly show
that the best models have $Z=0.012$. The effects of the other
parameters are not as striking as in the case of metallicity.
Yet, panels \ref{ddobscen1}(c) and \ref{ddobscen1}(d) 
favour an age in the range $8.75 \leq \log(\tau) \leq 8.80$.
Likewise, the three lowest values of intrinsic distance modulus 
(two of which are not even shown in the figure)
can essentially be ruled out (panels \ref{ddobscen2}(a) 
and \ref{ddobscen2}(b)), placing NGC\,1831 near or beyond 
the distance to the LMC centre.
Finally, $E(B-V) \leq 0.02$ is favoured by our modelling approach
(panels \ref{ddobscen2}(c) and \ref{ddobscen2}(d)).
Notice that this latter parameter is again anti-correlated with $Z$,
in the sense
that the models in the blowup panels with $Z = 0.008$ (circles in
panel \ref{ddobscen1}(b)) have larger $E(B-V)$ values (circles or crosses 
in panel \ref{ddobscen2}(d)).

These best choices of the global parameters, as discussed
in Sect. 3.1, besides being useful constraints by themselves,
serve as model inputs to the position dependent parameters $\alpha$ and
$f_\mathrm{bin}$, whose modelling
is based on the concentric regions. 

\subsection{Concentric regions}
\begin{figure} 
\resizebox{\hsize}{!}{\includegraphics{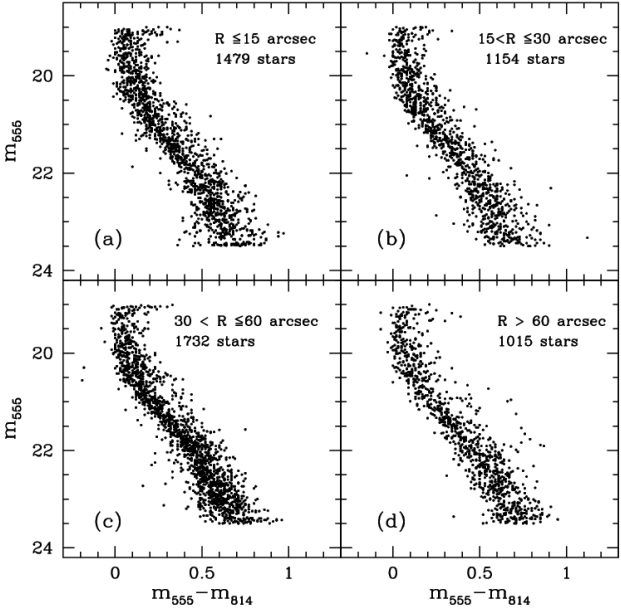}}
\caption[]{The final NGC\,1831 CMDs in four concentric regions:
(a) $0 < R \leq 15$ arcsec;
(b) $15 < R \leq 30$ arcsec;(c) $30 < R \leq 60$ arcsec;(d) $R > 60$ arcsec.}
\label{4cmd_rings}
\end{figure}

We now analyze the NGC\,1831 CMDs in the concentric regions 
listed in Table \ref{tab1}. As just mentioned, the free model parameters 
in this case are the PDMF slope $\alpha$ and the unresolved
binary fraction $f_\mathrm{bin}$.
For all regions we corrected the observed CMD to the effects
of sample incompleteness and field star contamination. Fig. \ref{4cmd_rings}
shows their final CMDs. As mentioned before, Table \ref{tab1} shows the number
of points in all the steps along the data correction process.

In the modelling, we apply the same model grid for every region, 
only changing the number of artificial stars generated. The 
results from the previous section were used to restrict
the possible values of the cluster´s global parameters, 
allowing us to investigate the 
dynamically dependent parameters for the system.

\begin{figure} 
\resizebox{\hsize}{!}{\includegraphics{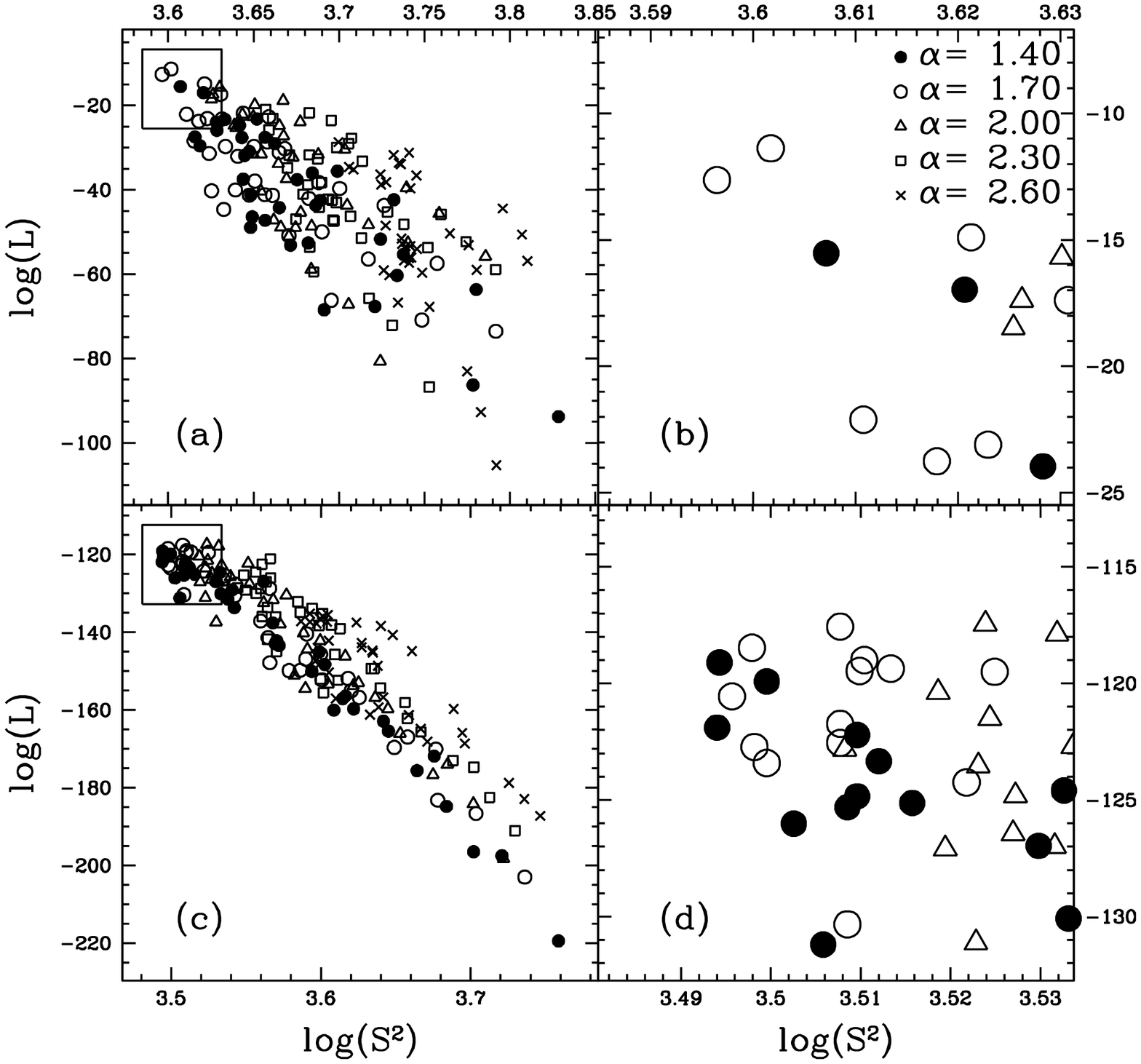}}
\caption[]{DDs resulting from the CMDs of concentric regions, 
showing the effect of varying $\alpha$. The upper (lower) 
panels show the results for stars in the region with $ 0 \leq R \leq 15$ 
($15 < R \leq 30$) arcsec. 
The panels on the left show the entire DD, while those on the right
show the best models in detail.
The symbols are as indicated in the panels on the right.}
\label{ddobsrings12}
\end{figure}

\begin{figure} 
\resizebox{\hsize}{!}{\includegraphics{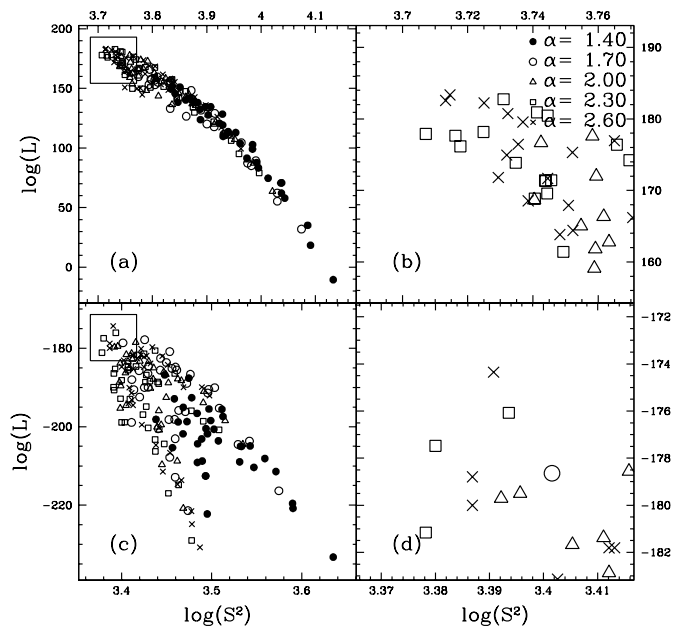}}
\caption[]{DDs resulting from the CMDs of concentric regions, 
showing the effect of varying $\alpha$. The upper (lower) 
panels show the results for stars in the region with $ 30 < R \leq 60$ 
($ R > 60$) arcsec. 
The panels on the left show the entire DD, while those on the right
show the best models in detail.
The symbols are as indicated in the panels on the right.}
\label{ddobsrings311}
\end{figure}

Figs. \ref{ddobsrings12} and \ref{ddobsrings311} present the 
DDs resulting from model vs. data CMD comparison in the four
regions. The different symbols in each panel correspond to
different values of $\alpha$. As usual, the panels on the right show the
best models in detail. The results for
the two innermost regions are presented in Fig. \ref{ddobsrings12}.
Panels (a) and (b) indicate that the best models for innermost region
($R \leq 15$ arcsec) have $1.40 \leq \alpha \leq 1.70$.
In the ring $15 < R \leq 30$ arcsec, 
shown in panels (c) and (d) of the same figure, the best models
present similar $\alpha$ values: $1.40 \leq \alpha \simlt 1.70$.

In Fig. \ref{ddobsrings311} we have the results for the two outermost rings. 
Panels (a) and (b) show the DDs for the stars with
$30 < R \leq 60$ arcsec. Now, there is evidence for a sharp
change in the PDMF´s slope: the best models have 
$2.30 \leq \alpha \leq 2.60$. Finally, the DD for the most peripheric ring 
has a large dispersion (panels \ref{ddobsrings311}(c) and 
\ref{ddobsrings311}(d) ). This spread is also seen 
in the best PDMF slope: values in the range 
$2.00 \simlt \alpha \leq 2.60$ are seen in the upper left corner of the DD.
This may reflect large uncertainties in the field star subtraction, since
 it leads to a large reduction of CMD points in this ring. 

For each ring, we assign a representative $\alpha$ and
associated uncertainty using the 10 models with the largest values of
likelihood $L$. The final slope is taken to be the weighted average value 
among these best models, and its associated uncertainty is the dispersion
around the average. The weight assigned to each model 
was the inverse of the difference in $\log L$ between the
observed CMD and a CMD from a typical model realization. This difference
is a measure of the discrepancy between model and observed CMDs. 
Table \ref{tab2} lists the final $\alpha$ values and uncertainties for each
ring (Cols. 3 and 4). The first 2 columns in the table
show the regions in arcsec and parsecs (assuming $(m-M)_{0}=18.61$
for NGC\,1831 as adopted in Sect. 6), respectively. 

\begin{table}
\caption[]{Best estimates of the slope $\alpha$ and its uncertainty as a 
function of distance to the centre.}
\label{tab2}
\small
\renewcommand{\tabcolsep}{1.1mm}
\begin{tabular}{lcccc}
\hline\hline
{Region} (arcsec)& Region (pc) & $\alpha$ & $\sigma_{\alpha}$ 
\\ \hline
\hline
$0 < R \leq 15$ & $0 < R \leq 3.8$ & 1.72 & 0.15 \cr 
$15 < R \leq 30$ & $3.8 < R \leq 7.6$ & 1.68 & 0.19 \cr
$30 < R \leq 60$ & $7.6 < R \leq 15.2$ & 2.45 & 0.15 \cr
$R > 60$ & $R > 15.2$ & 2.19 & 0.33 \cr
\hline\hline
\end{tabular}
\end{table}

As for the fraction of unresolved binaries, $f_\mathrm{bin}$, the results, as
anticipated, are not conclusive in any ring, given the CMD spread. 
A more precise treatment of photometric errors may help constrain this
parameter. 

\section{Summary and Conclusions}
In this work we analyzed a deep CMD of NGC\,1831, 
a rich LMC cluster,
obtained from HST/WFPC2 images in the F555W and F814W filters. 
We inferred physical parameters such as metallicity, age, 
intrinsic distance modulus,
reddening and PDMF slopes by comparing the cluster CMD with artificial
ones. We presented in detail the techniques used to build the model CMDs,
which use these parameters as model input and take into account observational
uncertainties as in the real data.
The parameter space explored by our regular model grids bracketed
the values found in the literature. 

The model vs. data comparison required correcting the observed CMD
for selection effects caused by photometric incompleteness and CMD 
contamination by stars belonging to the LMC field.
We also presented the statistical techniques used to compare the real CMD,
corrected for the aforementioned selection effects, to the
artificial ones. These statistical tools allowed us to discriminate 
the models that best 
reproduce the data. They are based on simple and objective 
statistical quantities and
make use of the full bidimensional distribution of points in the CMD.

The best parameter values inferred for NGC\,1831 are in the ranges
$8.75 \leq \log(\tau) \leq 8.80$, $18.50 \leq (m-M)_{0} \leq 18.70$, 
$0.00 \leq E(B-V) \leq 0.02$. Using the weighted average value among the
10 best models, as described in Sect. 5.2, we have
$\tau = 588 \pm 43$ Myr, $(m-M)_{0} = 18.61 \pm 0.07$,
$E(B-V) = 0.013 \pm 0.015$. As for the metallicity, all 10 best models
have $Z = 0.012$.

As discussed in Sect. 4, there is a coupling between reddening and 
metallicity, and therefore the determination of the former limits the 
values of the latter. As a consequence, the derived high 
metallicity for NGC\,1831 brings about a low reddening determination.

The combined values of the global parameters obtained in this work 
for NGC\,1831 suggest that this cluster is:
a) metal-richer and older than in most previous estimates;
b) placed near or beyond the distance to the LMC centre;
c) found in a low reddening region. 

Our estimated age and metallicity for NGC\,1831 are in 
perfect accordance with the age-metallicity relation
for LMC clusters obtained by Olszewski et al. (1991).

The best models for the global parameters were then used to build artificial 
CMDs of regions separated according to distance from the cluster centre
and with varying values of the PDMF slope and fraction of unresolved binaries.
For the PDMF slope $\alpha$ our statistical modelling shows that significant
mass segregation exists in NGC\,1831: for the central regions (out to
30 arcsec $\simeq 7.6$ pc), we derive $\alpha \simeq 1.70$, 
whereas $\alpha \simgt 2.20$ for the outer regions.
As for binarism, our results were not as conclusive. One explanation is that
the spread in the CMD caused by unresolved binaries is of similar or smaller
amplitude than the empirically derived photometric uncertainties in the data.
Therefore, this parameter is highly sensitive to the adopted photometric
errors, rendering its realistic estimate a task for yet deeper CMDs or
for data for which photometric uncertainties may be better
estimated and modelled.

In a preliminary analysis of these data, Santiago et al. (2001) 
observed the effect of mass segregation in NGC\,1831 with
a steepening of the LF slope as a function of distance from the
centre. However, no PDMF was derived. 
A global PDMF slope for NGC\,1831 is presented by Mateo (1988),
through the conversion of the luminosity function down to $V \sim 23$ 
($M \simgt \Msun$)
into a PDMF. The resulting slope is $\alpha \sim 4.0$, therefore
considerably steeper than our position dependent ones. Global slope
values of other clusters from Mateo (1988) are in 
the range $2.5 \simlt \alpha \simlt 4.6$.
In contrast, Elson et al. (1989)
find much shallower slopes $0.8 \simlt \alpha \simlt 1.8$ for another
sample of rich LMC clusters. 
These earlier results
are based on ground-based observations, for which the effects of crowding 
are much more serious than in the present work.

Our determined position dependent PDMFs constrain the 
$\alpha$ values within $\sim 0.2$. These results can be used
along with N-body simulations in order to recover the initial conditions,
in particular the cluster IMF.
We are applying the techniques shown in this paper to the others LMC clusters
imaged with HST/WFPC2 as part of the GO7307 project in order to infer
the same physical information as for NGC\,1831.
These future results can be very useful in investigations on the
IMF universality. 

\begin{acknowledgements}
We acknowledge CNPq and PRONEX/FINEP\,
76.97.1003.00 for partially supporting this work. 
We are grateful to R. de Grijs, S. Beaulieu, R. Johnson, G. Gilmore for
useful discussions. BXS is, as always, in debt with the late Becky Elson
for all he learned from her.
\end{acknowledgements}

%
%

\end{document}